\documentclass{PoS}

\def\fm {\mathop{\hbox{fm}}}
\def\MeV {\mathop{\hbox{MeV}}}

\def\solution#1  {}

\def\Tr {\mathop{\hbox{tr}}\nolimits}

\usepackage{amsmath} 
\usepackage{dsfont}  
\usepackage{bm}      

\def\beq{\begin{equation}}
\def\eeq{\end{equation}}

\def\comment#1{}

\def\opbraket#1#2#3{ \left\langle #1 \mid #2 \mid #3 \right\rangle}

\usepackage{graphicx}
\graphicspath{{figures/}}

\title{Absolute X-distribution and self-duality}

\ShortTitle{Absolute X-distribution and self-duality}

\author{\speaker{Andrei Alexandru}\\ 
        The George Washington University, Washington, DC, USA\\
        E-mail: \email{aalexan@gwu.edu}}

\author{Ivan Horv\'ath\\
        University of Kentucky, Lexington, KY, USA\\
        E-mail: \email{horvath@pa.uky.edu}}

\abstract{
Various models of QCD vacuum predict that it is dominated by excitations
that are predominantly self-dual or anti-self-dual. In this work we look at
the tendency for self-duality in the case of pure-glue SU(3) gauge theory
using the overlap-based definition of the field-strength tensor. To gauge
this property, we use the absolute X-distribution method which is
designed to quantify the {\em dynamical} tendency for polarization for
arbitrary random variables that can be decomposed in a pair of orthogonal
subspaces.
}

\FullConference{XXIX International Symposium on Lattice Field Theory \\
		 July 10-16 2011\\
		 Squaw Valley, Lake Tahoe, California}

\begin{document}

\section{Motivation}

Various models of QCD vacuum use semi-classical arguments to describe the
mechanism responsible for confinement or chiral-symmetry breaking. The
semi-classical arguments start by expanding QCD partition function around
extremal points of the action, i.e.,
\beq
\opbraket{\Omega}{e^{-H\tau}}{\Omega} \approx e^{-S_\text{cl}}\int{\cal D}x(\tau)\,
\exp \left( -\frac12\delta x\left.\frac{\delta^2S}{\delta x^2}\right|_{x_\text{cl}}
\delta x+\cdots \right) \,.
\eeq
The first task is then to find the extremal points of the action and then take
into account gaussian fluctuations around these extrema.

The action for pure-glue QCD can be expressed in terms of the self-dual and 
anti-self-dual components of the field strength tensor
\beq
S = \frac1{4g^2}\int d^4x\, F_{\mu\nu}^a F_{\mu\nu}^a = 
\frac1{4g^2} \int d^4x\, \left[\pm F_{\mu\nu}^a \tilde F_{\mu\nu}^a
+\frac12 \left(F_{\mu\nu}^a \mp \tilde F_{\mu\nu}^a \right)^2 \right] \,.
\eeq
The integral of the $F_{\mu\nu}^a \tilde F_{\mu\nu}^a$ term is a boundary term 
that is related to the
topological charge of the configuration. If we keep the boundary values fixed,
the integral is minimized when the quantity in the parenthesis vanishes.
This happens when the field is self-dual, $F_{\mu\nu}^a = \tilde F_{\mu\nu}^a$, or
anti-self-dual $F_{\mu\nu}^a = -\tilde F_{\mu\nu}^a$.
A more sophisticated analysis leads to the conclusion that all the extremal points 
of the classical action that are not saddle points satisfy 
this condition~\cite{Schafer:1996wv}.

It is then natural to expect that if QCD vacuum is correctly described by a 
semi-classical model, the field strength in a typical lattice QCD ensemble
will exhibit a high degree of  {\em self-duality}.
To gauge this tendency we decompose the
field strength at every point on the lattice into its self-dual components
and analyze their polarization properties.
To do this, we use the method of absolute X-distribution designed to analyze 
the {\em dynamical} aspects of polarization~\cite{Alexandru:2010sv}.
A more detailed account of this work is given in Ref.~\cite{Alexandru:2011yy}.

\section{Dynamical polarization}

We start by reviewing the method of absolute X-distribution. A first version
of this approach was introduced in a study of the local chirality of the low-lying 
eigenmodes of the Dirac operator~\cite{Horvath:2001ir}. In general, for an arbitrary observable 
that can be split in two components $Q=Q_1+Q_2$, we say that $Q$ is polarized when
it tends to be aligned with either one of the components. More precisely, if we 
look at the magnitude of components, $q_i = \Vert Q_i\Vert$, we tend to think that the
observable $Q$
is polarized when the probability distribution ${\cal P}_b(q_1,q_2)$, with support in the 
positive quadrant of the $q_1q_2$-plane, is peaked in the vicinity of the $q_{1,2}$ axes.

\begin{figure}[t]
\hbox to \hsize{
\includegraphics[width=2.5in]{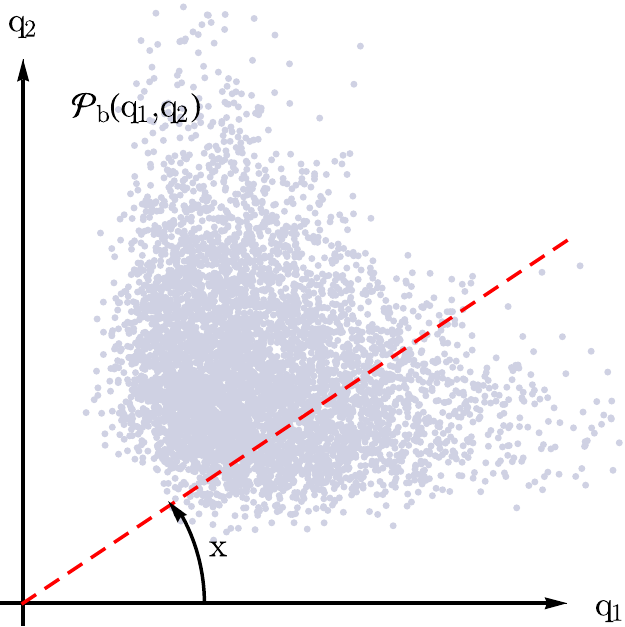} 
\hfill
\includegraphics[width=3in]{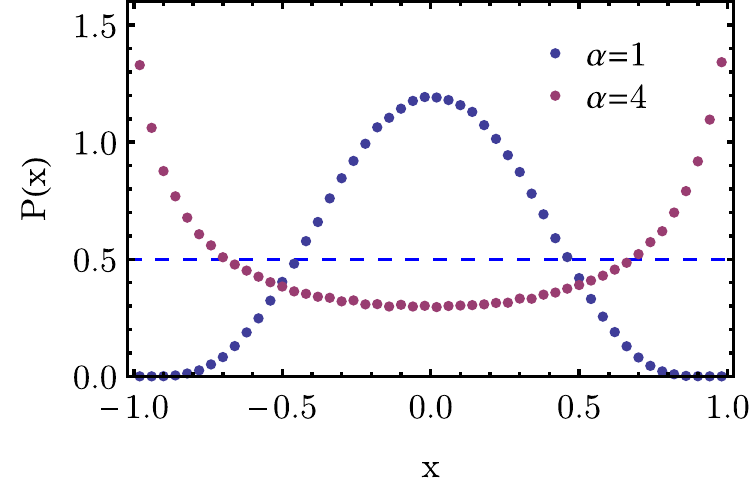} 
}
\caption{Sample pair distribution generated by chirality components of the lowest
eigenmodes of ensemble $E_1$ from~\cite{Alexandru:2010sv}. Right: the associated 
X-distribution, i.e., the induced 
distribution of the polarization angle.}
\label{fig:1}
\end{figure}

The raw distribution ${\cal P}_b(q_1,q_2)$ is difficult to characterize. A more direct measure
is offered by the induced distribution of the {\em polarization angle}. In
Fig.~\ref{fig:1} we plot the raw distribution of chirality components as determined in
a previous study~\cite{Alexandru:2010sv} and the corresponding polarization angle 
distribution (the curve indicated by $\alpha=1$), which we call the {\em X-distribution}. 
We see that the X-distribution 
tends to be concentrated towards the middle
of the graph, suggesting an anti-polarization tendency.

A more careful analysis reveals that the conclusions based on this method can be 
misleading. The X-distribution is determined by the choice of 
parametrization for the angles measured in the $q_1q_2$-plane. The definition
we used to plot Fig.~\ref{fig:1} is
\beq
x = \frac4\pi \arctan \frac{\Vert Q_2 \Vert}{\Vert Q_1 \Vert} - 1 \,.
\eeq
We will refer to this choice as the {\em reference polarization}~\cite{Horvath:2001ir}. 
However, this
choice is not unique. Alternative definitions were used in various studies. 
Using $t\equiv {\Vert Q_2 \Vert}/{\Vert Q_1 \Vert}$, one class of valid
angle variables is given by a generalization of the above definition
\beq
\bar{x} = \frac4\pi \arctan (t^\alpha) -1 \,,
\label{eq:2.2}
\eeq
where $\alpha>0$ is an arbitrary parameter~\cite{Alexandru:2010sv}. For $\alpha=1$ the
angle parameter $\bar{x}$ is the reference polarization defined above, while the definition
based on $\bar{x}$ with $\alpha=2$ was used in a study of 
self-duality in pure gauge QCD~\cite{Gattringer:2002gn}. 
In the right panel of Fig.~\ref{fig:1} we compare the X-distribution for
the ensemble shown in the left panel, measured using the reference polarization 
and the polarization defined by $\bar{x}$ with $\alpha=4$. The qualitative behavior 
of the distribution changes dramatically, while the dynamics producing the original
distribution is unchanged. It is clear then that conclusions based on X-distributions 
alone cannot be trusted.

To address this problem we define the {\em absolute X-distribution}, a measure of the
pair correlation induced by the underlying dynamics~\cite{Alexandru:2010sv,Draper:2004id}.
The basic idea is to compare the correlated distribution ${\cal P}_b(q_1,q_2)$ with
a similar distribution where the components are statistically independent, to isolate
the effect of the dynamics. 
The uncorrelated distribution is constructed from the marginal distributions
\beq
P_1(q_1) = \int dq_2\, {\cal P}_b(q_1,q_2) \qquad\text{and}\qquad
P_2(q_2) = \int dq_1\, {\cal P}_b(q_1,q_2)\,.
\eeq
For our application, symmetry guarantees that $P_1=P_2$. The uncorrelated
distribution is ${\cal P}_u(q_1,q_2)\equiv P_1(q_1) P_2(q_2)$. We define an 
angle variable that has constant angular density for the uncorrelated distribution.
This is the {\em absolute polarization}. The histogram of this angle variable for
the uncorrelated distribution is flat. In our figures this is indicated by 
a horizontal dashed line. The X-distribution in terms of the absolute polarization
is the {\em absolute X-distribution}.

\begin{figure}[t]
\hbox to \hsize{
\includegraphics[width=3in]{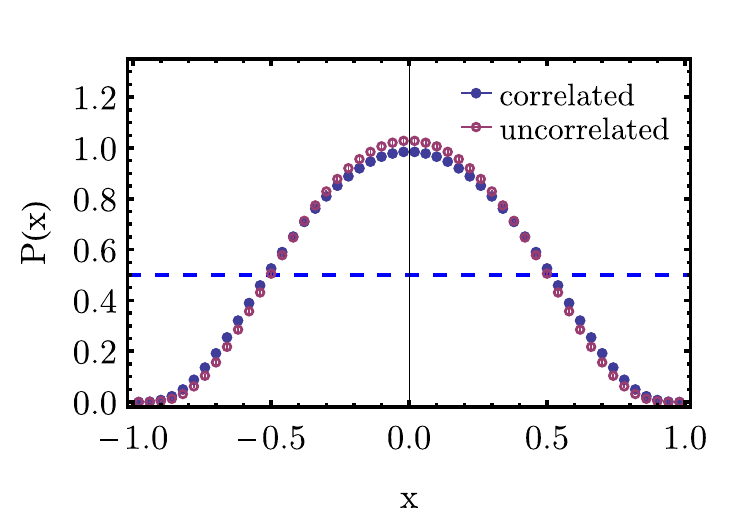} 
\hfill
\includegraphics[width=3in]{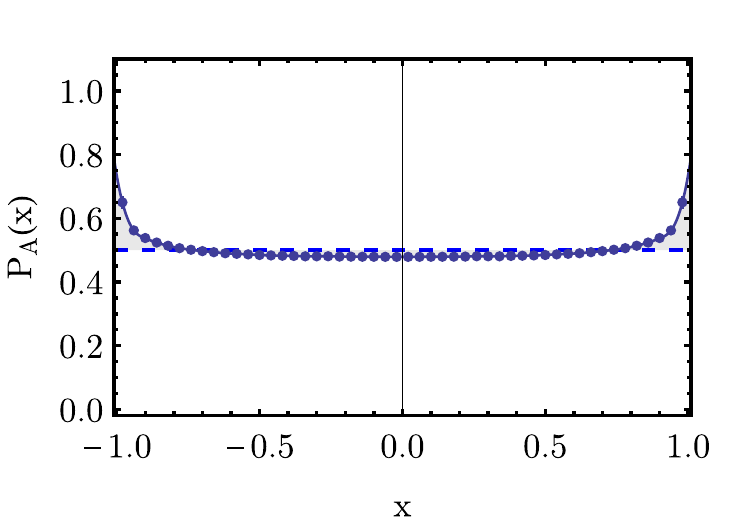} 
}
\caption{X-distribution using the {\em reference polarization} for the
correlated and uncorrelated distributions (left) and the {\em absolute X-distribution}
(right).}
\label{fig:2}
\end{figure}

In the left panel of Fig.~\ref{fig:2} we present the X-distribution for the
reference polarizations for both correlated distribution, ${\cal P}_b$,
and the uncorrelated one, ${\cal P}_u$, for the ensemble presented in Fig.~\ref{fig:1}. 
Notice that these two distributions are almost identical indicating that 
there is little dynamical correlation. In the right panel we plot the 
absolute polarization histogram, which is almost flat. There is a small enhancement 
towards the edges indicating that the dynamics induces a slight polarization.
This is consistent with the plots in the right panel, where
we see that the uncorrelated distribution is more prominent towards the
center of the histogram.

Based on the absolute polarization distribution, $P_A(x)$, we construct a more
compact measure of the polarization tendency, the {\em correlation coefficient}
\beq
C_A = 2\Gamma - 1 \qquad\text{where}\qquad \Gamma=\int\limits_{-1}^1 \!dx\,P_A(x) \left|x\right| \,.
\eeq
The coefficient $\Gamma$ measures the probability that a sample drawn from distribution
${\cal P}_b$ is more polarized than one drawn from ${\cal P}_u$. When we have
no dynamical correlation this probability is $0.5$; 
the correlation coefficient is scaled such that $C_A=0$ in this case.

\section{Field strength definition}
\label{sec:field-strength}

In this study, we will use a definition of the field strength based on the overlap 
operator. Compared to the ultra-local definitions, the overlap definition is less
susceptible to ultra-violet fluctuations, so no arbitrary link smearing or cooling is
needed. Moreover, this definition provides a natural expansion in terms of
eigenmodes of the Dirac operator which allows us to define a smoothed version
of field strength tensor controlled by the value of the eigenvalue cutoff.

If we denote with $S_F = \bar\psi D(x,y)\psi$ the fermionic contribution to the action
in the overlap formulation, it is easy to show that $\Tr_s \sigma_{\mu\nu} D(x,x)$ has
the same quantum numbers as the field strength $F_{\mu\nu}$~\cite{Horvath:2006md}. 
Here $\Tr_s$ denotes the
trace over the spinor index. It was shown by explicit calculation that on smooth
fields in the limit $a\to0$ these definitions agree~\cite{Liu:2007hq,Alexandru:2008fu},
\beq
\Tr\nolimits_s \sigma_{\mu\nu} D(x,x) = c^T F_{\mu\nu}(x)+{\cal O}(a^4) \,.
\eeq
Above, $c^T$ is a constant that depends on the kernel used to define the overlap operator.
The lattice version of the field strength operator used in this study is
\beq
F^\text{ov}_{\mu\nu}(x) \equiv \frac1 {c^T} \Tr_s \sigma_{\mu\nu} D(x,x) 
= -\frac1{c^T} \Tr_s\sigma_{\mu\nu}[2\rho - D(x,x)]\,,
\eeq
where $2\rho$ is the largest eigenvalue of $D$, the eigenvalue associated with
the zero modes' partners. We used the fact that $\Tr_s\sigma_{\mu\nu}=0$ to cast 
the definition in a form useful for eigenmode expansion.

Using the expansion in terms of the eigenmodes of the Dirac operator, we define
the smoothed version of the field strength~\cite{Alexandru:2010sv}
\beq
F^\Lambda_{\mu\nu}(x) \equiv -\frac1{c^T} \sum_{\vert\lambda\vert<\Lambda a}\Tr_s\sigma_{\mu\nu} 
(2\rho-\lambda) \psi_\lambda(x) \psi_\lambda(x)^\dagger \,.
\eeq
This definition has the property that $\lim_{\Lambda\to\infty} F^\Lambda = F$ and that
the contribution of the largest eigenmodes is suppressed. The self-dual and anti-self-dual
parts of the field strength are defined using the dual of the field strength 
$\tilde F_{\mu,\nu}=\frac12\epsilon_{\mu\nu\alpha\beta}F_{\alpha\beta}$
\beq
F_S = \frac12 (F+\tilde F) \qquad F_A = \frac12 (F-\tilde F) \,.
\eeq

\section{Numerical results}
\label{sec:results}

For our study we used a set of pure-glue ensembles generated using Iwasaki 
action~\cite{Okamoto:1999hi}. The parameters for these ensembles are presented
in Table~\ref{tab:1}. To study the continuum limit we have a set of 5 ensembles
with the same volume. To determine the finite volume effects we also generated
one ensemble with a larger volume.

\begin{table}[b]
\begin{center}
\begin{tabular}{c c c c c }
\hline\hline
Ensemble &Size& Lattice spacing & Volume & Configurations\\ \hline
$E_2$ & $12^4$ & $0.110\fm$ & & 400 \\ 
$E_3$ & $16^4$ & $0.083\fm$ & & 200 \\ 
$E_8$ & $20^4$ & $0.066\fm$ & $(1.32\fm)^4$ & 80\\ 
$E_4$ & $24^4$ & $0.055\fm$ & &  40\\ 
$E_7$ & $32^4$ & $0.041\fm$ & &  20\\ 
\hline
$E_6$ & $32^4$ & $0.055\fm$ & $(1.76\fm)^4$ & 20\\
\hline\hline
\end{tabular}
\end{center}
\caption{The size and lattice spacing for the ensembles used in this study.}
\label{tab:1}
\end{table}

\begin{figure}[t]
\hbox to \hsize{
\includegraphics[width=3in]{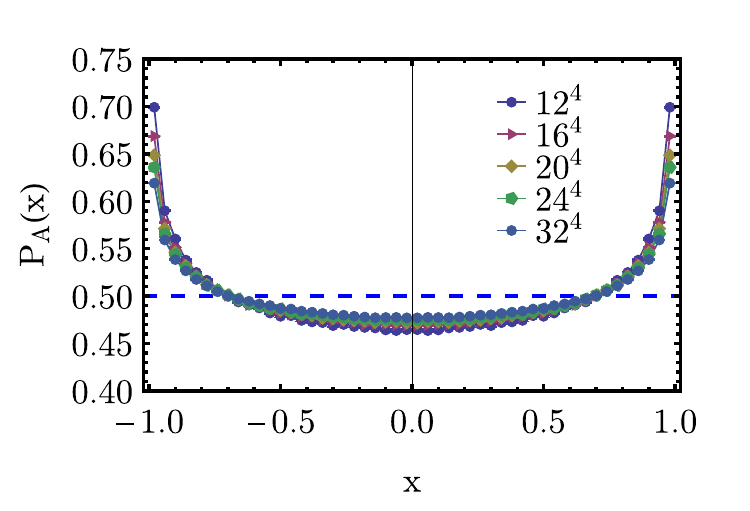} 
\hfill
\includegraphics[width=3in]{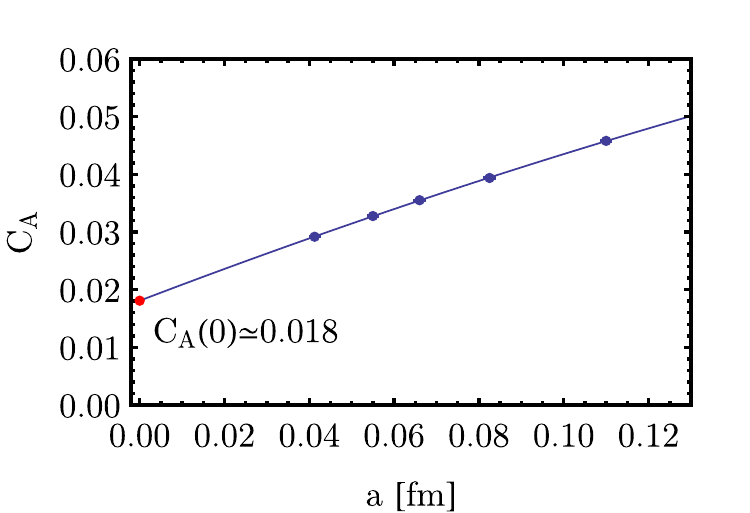} 
}
\caption{Left: absolute X-distribution for self-duality components. Note that the
y-scale is magnified to better show the difference between different lattice spacings.
Right: the correlation coefficient as a function of the lattice spacing and its
continuum limit extrapolation. Error bars are present in these plots but they are smaller
than the symbol size.}
\label{fig:3}
\end{figure}

In Fig.~\ref{fig:3} we plot the histogram for the absolute polarization for all ensembles
with volume $(1.32\fm)^4$. We find a small tendency for polarization that decreases as
we make the lattice spacing smaller. To understand whether this tendency survives the 
continuum limit, we compute the correlation coefficient and fit it with a quadratic 
polynomial in $a$. As we can see from the right panel of Fig.~\ref{fig:3} the polynomial
fits the data well. The coefficient remains positive in the continuum limit, indicating
a very small tendency for polarization. The probability that the sample
drawn from the correlated distribution is more polarized than one drawn from the
uncorrelated distribution is 51\% compared to 50\% when the dynamics would produce 
no correlation.

\begin{figure}[b]
\hbox to \hsize{
\includegraphics[width=3in]{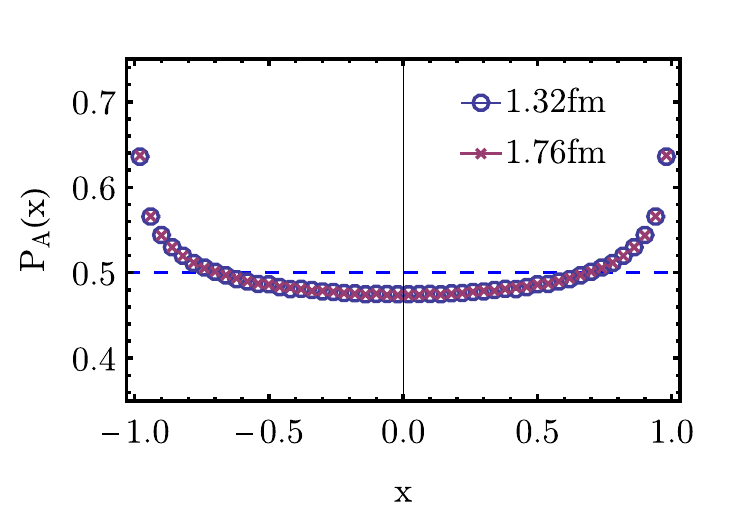} 
\hfill
\includegraphics[width=3in]{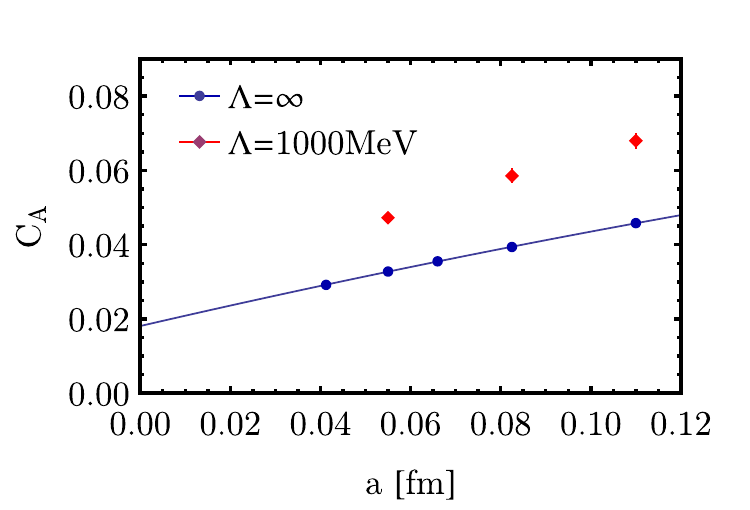} 
}
\caption{Left: absolute X-distribution for ensemble $E_4$ (circles) and $E_6$ (crosses) which 
have the same lattice spacing but different volume.
Right: correlation coefficient for the smoothed strength field (diamonds) compared to
the full version (circles). Error bars are included in both plots.}
\label{fig:4}
\end{figure}

To gauge the size of the finite volume effects, we compute the absolute polarization
on two ensembles with the same lattice spacings but different volumes. Referring
to Table~\ref{tab:1}, these are ensembles $E_4$ and $E_6$. In the left panel of
Fig.~\ref{fig:4} we compare the absolute polarizations on these two ensembles. We find
no difference between the two histograms and we conclude that the finite volume 
effects are negligible.

We also computed a set of eigenmodes of the overlap Dirac operators on ensembles
$E_2$, $E_3$ and $E_4$ and used them to compute the smoothed field strength operator
$F^\Lambda$. To study the continuum limit, a consistent definition 
of the smoothed operator sums over all modes smaller than a physical cutoff. We
set the cutoff $\Lambda = 1000\MeV$ and found that the behavior of the absolute 
X-distribution is similar to the full version of the operator. In the right
panel of Fig.~\ref{fig:4} we compare the correlation coefficient with the one
computed using the full operator. We find that while the values of the correlation
coefficient are slightly different, the qualitative behavior remains the same.

\begin{figure}[t]
\hbox to \hsize{
\includegraphics[width=2.23in]{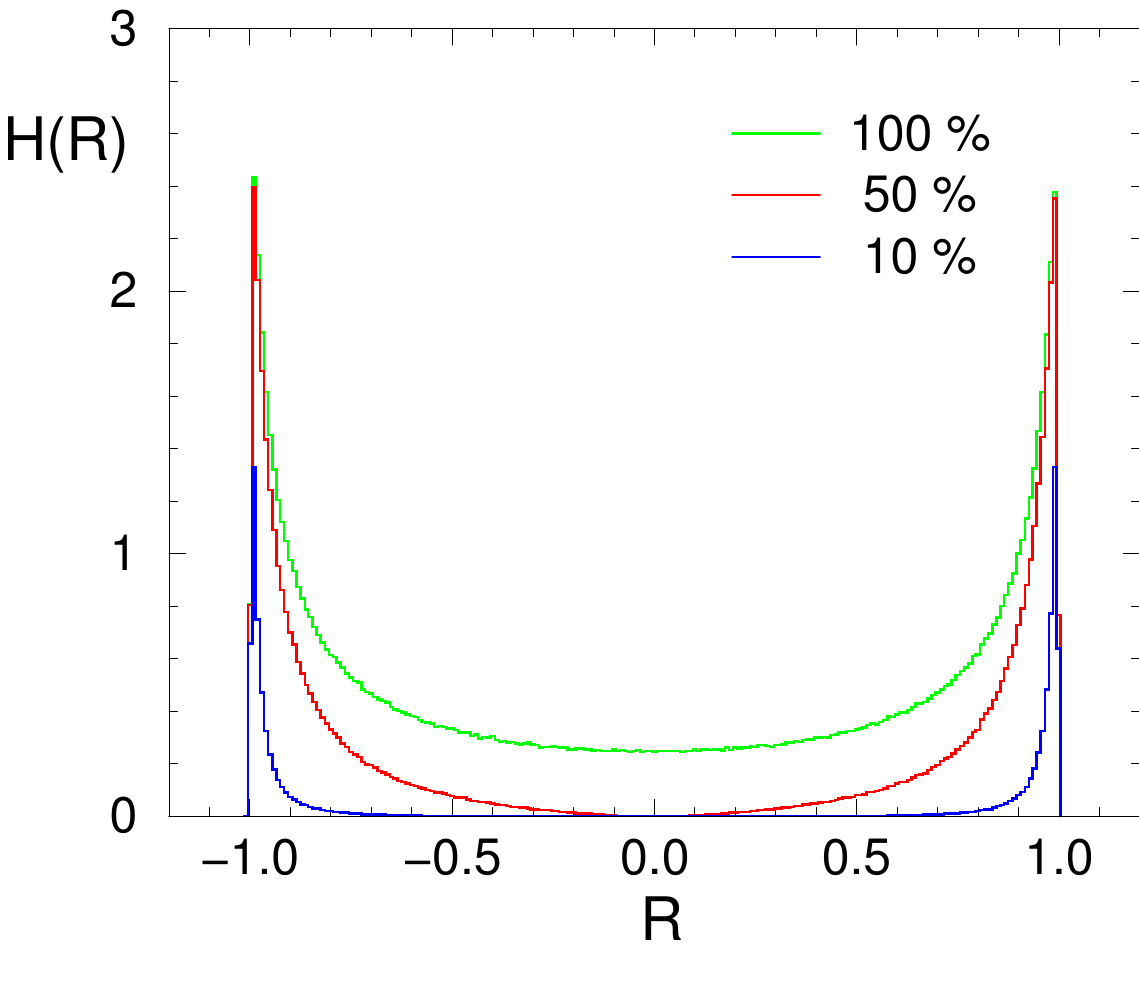} 
\hfill
\includegraphics[width=3in]{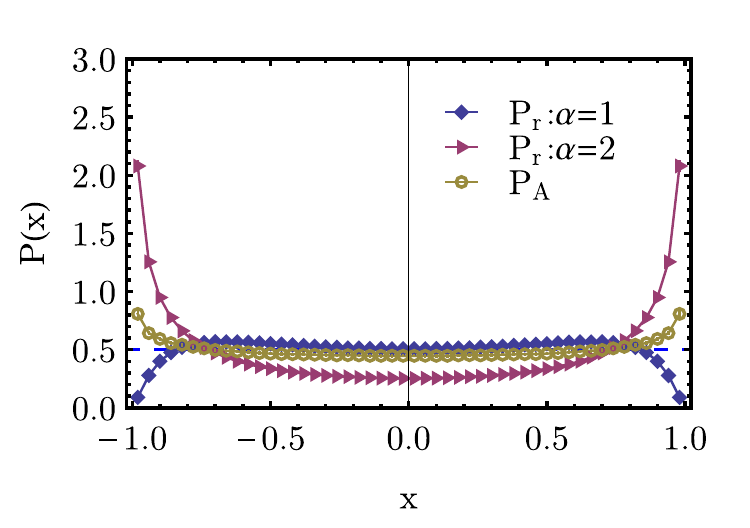} 
}
\caption{Left: X-distribution for self-duality components of a smooth field strength
based on the low-lying modes of the chirally-improved Dirac operator~\cite{Gattringer:2002gn}.
The curved marked with 100\% is the relevant one for our comparison.
Right: absolute X-distribution $P_A$ and X-distribution $P_r$
based on two different polarization variables (see {Eq.~\protect\ref{eq:2.2}})
for ensemble $E_2$.}
\label{fig:5}
\end{figure}

We conclude our discussion with a comparison with a similar work by Gattringer~\cite{Gattringer:2002gn}
who studied the self-duality polarization using a smoothed field strength 
operator. 
This operator was constructed using an eigenmode expansion of the chirally-improved Dirac 
operator. In Ref.~\cite{Gattringer:2002gn} it was found that the self-duality exhibits
a strong polarization (see left panel of Fig.~\ref{fig:5}) supporting a 
model of vacuum dominated by topological ``lumps". 
In contrast, we only find a mild dynamical tendency for polarization. 
This is seen in the right panel
of Fig.~\ref{fig:5} where we plot the absolute X-distribution of ensemble $E_2$ which is
similar to the ensemble used in Ref.~\cite{Gattringer:2002gn}. The discrepancy is due 
to the fact that Ref.~\cite{Gattringer:2002gn} uses a polarization measure dominated 
by kinematical effects. To show this, in the right panel of Fig.~\ref{fig:5} 
we also plot the X-distribution measured
using the reference polarization, $\alpha=1$, and the polarization angle
used in Ref.~\cite{Gattringer:2002gn}, $\alpha=2$. To better compare our results, for these
plots we used, as in the referenced study, a smoothed $F^\Lambda$ constructed using the same number of modes. We see then that when using the same angle
definition, our results are consistent with those of Ref.~\cite{Gattringer:2002gn}.
However, using another valid angle parametrization produces qualitatively different 
results due to kinematical effects.  
We conclude that the strong polarization observed in 
Ref.~\cite{Gattringer:2002gn} is mainly due to the specific choice of angle 
variable rather than the underlying
dynamics.

\section{Conclusions}

In this work we studied the dynamical polarization propertied of self-duality components 
induced by pure-glue QCD dynamics. We found a very mild polarization tendency that
survives in the continuum limit. This results has negligible finite-volume corrections. 
The self-duality tendency is very small making it unlikely that
the vacuum fluctuations are well-described by semi-classical models.
Our findings are at variance with the results of a previous study~\cite{Gattringer:2002gn}.
We conclude that the discrepancy is the result of kinematical effects. 

\smallskip
\noindent{\bf Acknowledgments}: Andrei Alexandru is supported in part under DOE 
grant DE-FG02-95ER-40907. The computational resources for this project were provided 
in part by the George Washington University IMPACT initiative. Ivan Horv\'ath acknowledges 
warm hospitality of the BNL Theory Group during which part of this work has been completed.

\bibliographystyle{JHEP}
\bibliography{my-references}

\end{document}